\makeatletter

\newcommand{\Rmnum}[1]{\expandafter\@slowromancap\romannumeral #1@}
\makeatother

\documentclass[conference]{IEEEtran}

\usepackage{multirow}
\usepackage[psamsfonts]{amssymb}
\usepackage{amsmath}
\usepackage{amsxtra}
\usepackage{amsmath, amsthm, amssymb}
\usepackage{threeparttable}
\usepackage{graphicx}
\usepackage{verbatim}   
\usepackage{subfigure}  
\usepackage{amsmath}    
\usepackage{makeidx}  
\usepackage{slashbox}  
\usepackage{times}
\usepackage{subfigure}
\usepackage{algorithm}
\usepackage{soul}
\usepackage{color} 
\IEEEoverridecommandlockouts
\definecolor{Light}{gray}{.90}

\ifCLASSINFOpdf
\else
\fi
\usepackage{algorithmic}

\begin{document}
%
\title{On Transmit Antenna Selection for Multiuser MIMO Systems with Dirty Paper Coding}


\author{\IEEEauthorblockN{Manar Mohaisen, \textit{Student Member, IEEE},  and KyungHi Chang, \textit{Senior Member, IEEE}}
\IEEEauthorblockN{The Graduate School of IT and T, Inha University\\
253 Yonghyun-Dong, Nam-Gu, 402-751 Incheon, KOREA\\
Email: lemanar@hotmail.com, khchang@inha.ac.kr}
\thanks{This work was supported by the Korea Science and Engineering Foundation (KOSEF) grant funded by the Korea government (MOST) (No. R01-2008-000-20333-0).}}

%


\maketitle

\begin{abstract}
In this paper, we address the transmit antenna selection in multi-user MIMO systems with precoding. The optimum and reduced complexity sub-optimum antenna selection algorithms are introduced. QR-decomposition (QRD) based antenna selection is investigated and the reason behind its sub-optimality is analytically derived. We introduce the conventional QRD-based algorithm and propose an efficient QRD-based transmit antenna scheme (maxR) that is both implementation and performance efficient. Moreover, we derive explicit formulae for the computational complexities of the aforementioned algorithms. Simulation results and analysis demonstrate that the proposed maxR algorithm requires only 1$\%$ of the computational efforts required by the optimal algorithm for a degradation of 1dB and 0.1dB in the case of linear zero-forcing and Tomlinson-Harashima precoding schemes,  respectively.

\end{abstract}


%
\IEEEpeerreviewmaketitle

\section{Introduction}
Dirty paper coding (DPC), first proposed by Costa \cite{Costa}, is the optimal capacity precoding approach for downlink multi-user multiple-input multiple-output (MU-MIMO) communication systems. DPC schemes use the channel state information (CSI), available at the transmitter by means of feedback, and the a priori known users' data to eliminate or reduce the co-channel interference (CCI). Therefore, the number of available transmit antennas at the base station (BS) should be larger than or equal to the total number of receive antennas at the mobile stations (MSs).\\
\indent Linear precoding schemes have low computational complexity but they are susceptible to amplify the noise or to introduce inter-user interference \cite{Peel}. Tomlinson-Harashima precoding (THP) is a nonlinear scheme that overcomes the noise amplification by introducing the modulo (MOD) operation, which reduces the required transmit power \cite{Tomlinson}, \cite{Harashima}. The error performance of the aforementioned precoding schemes is still far to fulfill the requirements of the future generations of mobile communication systems.\\
\indent Transmit antenna selection remarkably improves the system performance by exploiting the spatial selectivity. Therefore, when the number of antennas available at the BS is larger than the number of RF chains, the subset of antennas with the best channel conditions can be selected and switched to the RF chains.\\
\indent Antenna selection is broadly researched for single user MIMO multiplexing systems (\cite{Gucluoglu}-\cite{Sanayei2} and references therein), but fewer results were obtained for MU-MIMO systems \cite{Chen}. In this paper, we address the transmit antenna selection for MU-MIMO systems with precoding.\\
\textbf{Contributions.} Our original contributions in this paper are as following.
\begin{itemize}
	\item We introduce the optimal selection algorithm for DPC MU-MIMO system and a suboptimal reduced complexity algorithm that achieves a quasi-optimal performance.
	\item We investigate the reason behind the sub-optimality of the QR-decomposition-based (QRD-based) antenna selection. Then, we introduce a reduced complexity and low latency QRD-based antenna selection algorithm that achieves optimum diversity and lags the optimum performance by less than 1dB and 0.1dB in case of linear precoding and THP, respectively.
	\item Finally, we derive explicit formulas for the computational complexities of the introduced algorithms as a function of the number of required complex operations. 
\end{itemize}
The remaining parts of this paper are as following. Section II introduces the system model and a review of the precoding schemes. In section III, we introduce the optimum SNR-based selection algorithm and its reduced complexity version . In Section IV, QRD-based antenna selection is investigated, and the proposed maxR selection algorithm is introduced. The complexities of antenna selection algorithms is derived in Section V, simulation results are shown in Section VI, and conclusions are drawn in Section VII.
\section{System Model and Review of Precoding Techniques}
\subsection{System Model}
We consider a down-link (DL) MU-MIMO system, where a BS communicates simultaneously with $K$ non-cooperative MSs. Also, BS is equipped with $M$ RF chains and $N > M$ antennas, and each MS has a single antenna. In this paper, we consider the number of RF chains is equal to the number of MSs, i.e., $M = K$, and the best $M$ out of $N$ antennas are selected for the DL communication with the MSs. Under the assumption of narrow-band flat-fading channel, the MU-MIMO system can be modeled as following:
\begin{equation}
\textbf{y} = \textbf{Hx} + \textbf{n},
\end{equation}
where $\textbf{y} \in \mathbb{C}^{M}$ is the vector whose element $y_k$ is the received signal at the $k$-th MS, and $\textbf{x} \in \mathbb{C}^{M}$ is the precoded transmitted vector. $\textbf{n} \in \mathbb{C}^{M}$ is the additive white Gaussian noise whose mean and variance are zero and $\sigma_n^2$, respectively. Finally, \textbf{H} is the channel matrix whose element $h_{k, i}$ is the transfer function between the $i$-th transmit antenna and the single antenna of the $k$-th MS. The elements of \textbf{H} are independent and follow complex Gaussian distributions.
\subsection{Linear Precoding Schemes}
Linear zero-forcing precoding (LZF) cancels the effect of the 
channel by precoding the transmitted data vector using the pseudo-inverse of the channel 
matrix.
\begin{equation}
\textbf{x}_{\text{zf}} = \frac{1}{\sqrt{\gamma}} \textbf{H}^{\dagger} \textbf{s},
\end{equation}
where the scaling factor ${\gamma}$ is present to fix the \textit{expected} total transmit power to ($P_T$); that is,
\begin{equation}
\gamma = \frac{1}{\text{P}_T} \mathrm{Tr}\left\{\left(\textbf{HH}^H\right)^{-1}\right\},
\end{equation}
where $\mathrm{Tr}(\cdot)$ refer to the trace operation. As a consequence, the receive SNR at any MS is given by:
\begin{eqnarray}
 \mbox{SNR} = \frac{\mbox{E}(ss^*)}{\gamma \sigma_n^2}.
\end{eqnarray}
If the channel matrix is ill-conditioned, $\gamma$ becomes large and consequently the post-processing signal to noise ratio (SNR) is decreased. To overcome this drawback, linear minimum mean-square error (MMSE) precoding  can be used to regularize the channel matrix. The precoded signal using LMMSE is given by:
\begin{equation}
\textbf{x}_{\text{mmse}} = \textbf{H}^H\left(\textbf{HH}^H + \alpha \textbf{I}_{Nr}\right)^{-1}\textbf{s},
\end{equation}
where $\alpha$ = $K\sigma_n^2/P_T$ is the regularization factor.
\begin{figure}[!t]
\centering
\includegraphics[width=8cm]{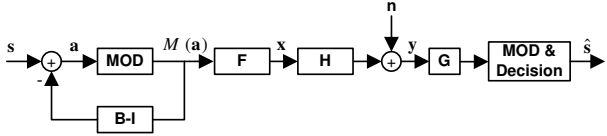}
\caption{Block diagram of Tomlinson-Harashima precoding.}
\label{THP_BD}
\end{figure}
\subsection{Tomlinson-Harashima Precoding (THP)}
In the case of ill-conditioned channel matrix, linear precoding leads to degradation in the receive SNR. To overcome this problem, THP includes the non-linear modulo operation (MOD) as shown in Figure \ref{THP_BD}. The modulo operation is defined by:
\begin{equation}
M(a) = a - \left\lfloor \frac{a}{{\tau}} + (\frac{1}{2} + j\frac{1}{2})\right\rfloor {\tau},
\end{equation}
where $\tau$ is chosen depending on the modulation constellation, and $\lfloor b \rfloor$ rounds the real and imaginary parts of $b$ to the closest lower integers. The feedforward (\textbf{F}) and feedback (\textbf{B}) matrices are obtained using the QR-decomposition of the Hermitian transpose of $\textbf{H}$. Let $\textbf{H}^H = \textbf{QR}$, then $\textbf{F} = \textbf{Q}^H$, and $\textbf{B} = \textbf{G}\textbf{R}^H$. $\textbf{G} = diag\left(1/R_{1,1}, 1/R_{2,2}, \cdots, 1/R_{M,M}\right)$ is a diagonal matrix whose elements are used as scaling factors at the receivers.\\
\indent A better performance can be achieved when the channel matrix is regularized as following:
\begin{equation}
\tilde{\textbf{H}} = [\textbf{H}\;\;\;\; \sqrt{\alpha}\textbf{I}_M] \in \mathbb{C}^{M\times N+M}.
\end{equation}
Then the QRD of the matrix $\tilde{\textbf{H}}^H$ is applied to obtain the feedforward and feedback matrices.\\
\indent Due to space limitation, we will only consider throughout this paper LZF and ZF-THP algorithms.
\subsection{Achievable Sum Rates by Precoding Schemes}
For linear pre-coding, the post-processing signal to noise ratio ($\text{SNR}_p$) is inversely 
proportional to the scaling factor $\gamma$, i.e., $\text{SNR}_p = \rho/\gamma$, 
where $\rho = \text{E}(s^*s)/\sigma^2$. 
Therefore, the sum rate capacity of linear ZF precoding is given by:
\begin{equation}
C_{\text{zf}} = K\cdot \text{E}\left[\log_2(1 + \rho/\gamma)\right],
\end{equation}
where the maximum achievable sum rate of $K\cdot\log_2(1 + \rho)$ is attained 
when the antennas are perfectly uncorrelated.\\
\indent On the other hand, the sum rate for ZF-THP is given by:
\begin{equation}
C_{\text{zf-THP}} = \sum_{i=1}^{K}\log_2(1 + \rho R_{i, i}^2).
\end{equation}
\section{SNR-based Antenna Selection}
\subsection{SNR-based Optimum Antenna Selection}
When linear pre-coding algorithms are employed, the scaling factor $\gamma$ is fixed for all users. 
Thus, $\gamma$ is selected in such a way the post-precessing SNR is maximized. 
Thus, to optimize the post-processing SNR, the optimal antenna subset is selected as:
\begin{align}
{\cal{A}}_{opt} &= \underset{{\cal{A}} \in {\cal{S}} \,\,\, p = 1, \cdots, P}{\operatorname{arg\,min}} \, 		
												\gamma\left\{\textbf{H}_p\right\},
\label{eq:1}
\end{align}
where $P = C_M^N$, $\gamma\left\{\textbf{H}_p\right\} = \mathrm{Tr}\left\{\left(\textbf{H}_p\textbf{H}_p^H\right)^{-1}\right\}$, and $\text{P}_T =$ 1, i.e., $\text{SNR} = 1/\sigma_n^2$. Although this algorithm is optimized for linear precoding schemes, it is also performance optimum for successive interference cancellation because the best-conditioned matrix is selected.\\
\indent The optimum antenna subset ${\cal{A}}_{opt}$ is obtained by employing
 exhaustive search over the $C_M^N$ possible subsets. 
 
\subsection{SNR-based Reduced-Complexity (RC) Antenna Selection}
To reduce the complexity of the SER-based optimum antenna selection, we propose 
to successively remove antennas one-by-one until the remaining number of antennas 
equals to $M$. This strategy is adapted from \cite{Chen}, where antenna selection algorithm are investigated for MU-MIMO with block diagonalization. At each iteration, the antenna whose deactivation maximizes the post-processing SNR, i.e., minimizes $\gamma$, is removed. This procedure is repeated iteratively till the number of remaining antennas is equal to $M$. 
The full description of the reduced complexity antenna selection is described 
in Fig. \ref{rcAlg}. Note that $\textbf{H}_p = \textbf{H}(:, {\mathcal{A}}_{rc})$ 
is the sub-matrix of \textbf{H} with elements from rows 1 to $M$ and columns indexes $\in {\mathcal{A}}_{rc}$.\\
\indent Therefore, to obtain the subset ${\mathcal{A}}_{rc}$, only partial antenna 
combinations are to be searched over instead of the $C_{M}^N$ possible combinations. 

\begin{figure}[t]
    \begin{algorithmic}[1]
    \small{
		\REQUIRE \textbf{H}, $N$, $M$
			\STATE $U = N$, $V = N - M$, 	${\mathcal{A}}_{rc} = \mathcal{S}$ 	
				\FOR{$i = 1$ to $V$}
							\FOR{$k = 1$ to $U$}
									\STATE $\textbf{H}_s = \textbf{H}^{-k}$ \hspace{2cm}   \COMMENT{remove the $k^{th}$ column}
									\STATE $\gamma_k$ = $\mathrm{Tr}\left\{\left(\textbf{H}_s\textbf{H}_s^H\right)^{-1}\right\}$
							\ENDFOR
							\STATE $k_{i} =$ $\underset{{k = 1, \cdots, U}}{\operatorname{arg\,min}} \, \left(\gamma_k\right)$
							\STATE $U = U - 1$             
		\STATE ${\mathcal{A}}_{rc} = {\mathcal{A}}_{rc} - \left\{k_{i}\right\}$\hspace{1.8cm}\COMMENT{deactivate antenna $k_{i}$}
							\STATE $\textbf{H} = \textbf{H}^{-k_{i}}$ \hspace{2.4cm}   \COMMENT{remove the $k_{i}^{th}$ column}
			  \ENDFOR
			  \STATE $\textbf{H}_p = \textbf{H}(:, {\mathcal{A}}_{rc})$
		\ENSURE $\textbf{H}_p$, ${\mathcal{A}}_{rc}$}
		\end{algorithmic}
	\caption{SNR-based reduced-complexity antenna selection algorithm.}
	\label{rcAlg}
\end{figure} 
\section{QRD-based Antenna Selection}
The QR-decomposition (QRD) of the channel matrix was used in the literature to select  
a \textit{good but not certainly best} subset of the transmit antennas in 
spatial multiplexing systems. Although, the achieved performance of MIMO system 
employing QRD antenna selection is not optimum, it is preferable due to its low computational 
complexity.\\
\indent In light of the QRD of the channel matrix, the scaling factor is given as a function of the matrix \textbf{R} as follows:
\begin{align}
   \gamma &= \mathrm{Tr}\left\{\left(\textbf{HH}^H\right)^{-1}\right\}, \nonumber\\
					&= \mathrm{Tr}\left\{\left(\textbf{QR}\textbf{R}^H\textbf{Q}^H\right)^{-1}\right\},\nonumber\\
					&= \mathrm{Tr}\left\{\left(\textbf{R}^{-H}\textbf{R}^{-1}\right)\right\}.
\label{eq:qr1}
\end{align}
Let \textbf{A} = $\textbf{R}^{-1}$, then, due to the triangular structure of \textbf{R}
\begin{align}
\mathrm{Tr}\left\{\left(\textbf{R}^{-H}\textbf{R}^{-1}\right)\right\} 
							& = \sum_{i = 1}^{M}\sum_{j = i}^{M} \left\|A_{i,j}\right\|^2,\nonumber\\
							&= \sum_{i = 1}^{M} \frac{1}{R_{i, i}^2} + 
							\sum_{i = 1}^{M}\sum_{j = i+1}^{M} \left\|A_{i,j}\right\|^2,\nonumber\\
							&\geq \sum_{i = 1}^{M} \frac{1}{R_{i, i}^2},
							\label{eq:qr2}
\end{align}
where $R_{i,i}$ are the entries on the diagonal of \textbf{R}. The equality in (\ref{eq:qr2}) 
is satisfied iff \textbf{H} is orthogonal matrix, i.e., \textbf{R} is diagonal.\\
\indent Therefore, a sub-optimum antenna subset can be obtained by minimizing 
$\sum_{i = 1}^{M} \frac{1}{R_{i, i}^2}$. This can be done by maximizing the diagonal 
elements of \textbf{R}. For that, QRD-based antenna selection algorithms are not optimum due to discarding the second summation of (\ref{eq:qr2}). In the following, we introduce two QRD-based antenna selection algorithms.
\subsection{Antenna Selection using a Single QRD (Single-QR)}
In the literature, single-QR antenna selection algorithm was used for spatial 
multiplexing systems which employ successive interference cancellation detection. 
In this paper we employ  single-QR algorithm for transmit antenna selection in 
MU-MIMO system with precoding. The main idea behind single-QR algorithm is to maximize 
the $\sum_{i=1}^{M}R_{i,i}^2$ by using only one QR decomposition of the channel matrix. At the first iteration, the column of the channel with the largest power is selected. The remaining ($N-1$) columns are then orthogonalized, where the orthogonalized column with the largest power is considered for the second iteration. This process is repeated till selecting $M$ out of the $N$ possible transmit antennas.\\
\indent Although single-QR algorithm has respectively low complexity, 
the selection of the column with the largest power, which may have high correlation with other columns, at the first iteration can lead to degradation in its performance as compared to that of the optimum antenna selection algorithm.\\
\indent In the following Section we introduce a performance and complexity efficient 
antenna selection algorithm based on the QRD.
\begin{figure}
    \begin{algorithmic}[1]
    \small{
		\REQUIRE \textbf{H}, $N$, $M$
			\STATE $\textbf{R} = \textbf{0}_M$, $\textbf{Q} = \textbf{H}$, $bm = \infty$, ${\mathcal{A}}_{maxR} = \mathcal{S}$	
				\FOR{$i = 1$ to $N$}
						\STATE $\textbf{norms}_i$ = $\left\|\textbf{h}_i\right\|^2$
				\ENDFOR
				\STATE $\textbf{w} = $ sort$\left(\textbf{norms}_i, \text{descend}\right)$
				\FOR{$j = 1$ to $N$}
							\STATE $\textbf{Tnorms} = \textbf{norms}$, $\textbf{R} = \textbf{0}_M$
							\STATE $\textbf{Q} = \textbf{H}$, $D = 0$, ${\mathcal{P}} = {\mathcal{S}}$
							\FOR{$i = 1$ to $M$}
									\IF{$i = 1$}
										\STATE $k = w_j$
									\ELSE
										\STATE $k = \underset{m = i, \cdots, N}{\operatorname{arg\,max}}$$\left(\textbf{Tnorms}_m\right)$
									\ENDIF
									\STATE $D = D + 1/\textbf{Tnorms}_k$
									\IF{$D \geq bm$}
										\STATE break
									\ENDIF
									\STATE {Exchange columns $i$ and $k$ in \textbf{Q}, \textbf{R}, \textbf{Tnorms} and $\mathcal{P}$}
									\STATE $R_{i,i}$ = $\left|\textbf{q}_i\right|$
									\STATE $\textbf{q}_i$ = $\textbf{q}_i$/$R_{i,i}$
									\FOR{$m = i + 1$ to $N$}
										\STATE $R_{i,m}$ = $\textbf{q}_i^H \cdot \textbf{q}_m$
										\STATE $\textbf{q}_m$ = $\textbf{q}_m$ - $R_{i,m} \cdot \textbf{q}_i$
										\STATE $\textbf{Tnorms}_m$ = $\textbf{Tnorms}_m$ - $R_{i, m}^2$
									\ENDFOR
									\IF{$i = M$}
										\STATE ${\mathcal{A}}_{maxR} = \mathcal{P}(1:M)$
										\STATE $bm = D$
									\ENDIF
							\ENDFOR
			  \ENDFOR
			  \STATE $\textbf{H}_p = \textbf{H}(:, {\mathcal{A}}_{maxR})$
		\ENSURE $\textbf{H}_p$, ${\mathcal{A}}_{maxR}$}
		\end{algorithmic}
\caption{Maximum-R (maxR) Antenna Selection Algorithm.}
\label{maxAlg}
\end{figure}
\subsection{Maximum-R (maxR) Antenna Selection Algorithm}
To ensure obtaining the best antenna subset among the $C^N_M$ combinations, 
an exhaustive search should be employed to obtain the subset with minimum 
$\sum_{i=1}^M \frac{1}{R_{i,i}^2}$. Although this search scheme can be done 
successively, where unnecessary computations are avoided, its extreme-case complexity 
is still inapplicable.\\
\begin{table*}[!t]
\renewcommand{\arraystretch}{1.3}
\caption{Computational complexity of the transmit antenna selection schemes.}
\label{table_complexity}
\centering
\begin{tabular}{|c|p{8cm}|p{2.5cm}|p{2.5cm}|}
\hline
\textbf{Scheme} & \textbf{Complexity} (flops) & $\text{C}/{\text{C}_{\text{opt}}}\times 100 \newline (N = 8, M = 4)$& $\text{C}/{\text{C}_{\text{opt}}} \times 100 \newline (N = 14, M = 10)$\\
\hline
\textbf{Optimum} & $C_M^N \cdot \left(10{M}^{3}-\frac{1}{2}{M}^{2}+\frac{1}{2}M-1\right)$ &100 &100\\
\hline
\textbf{RC} &  $M^2(\frac{2}{3}N^3 + 4MN^2 + \frac{5}{3}M + 4MN -\frac{14}{3}M^3 - \frac{41}{12}M^2 - \frac{5}{4}N^2 - \frac{23}{12}N + \frac{11}{12}) + M(\frac{2}{3}N^3 + \frac{1}{4}N^2 - \frac{5}{12}N + \frac{1}{2}) - \frac{N^2 + N}{2}$& 41.1 & 5.2\\
\hline
\textbf{maxR} & $8N^2M^2 + \frac{1}{4}N^2 + \frac{3}{2}N^2M + \frac{11}{4}NM - 4NM^3 - \frac{15}{4}NM^2 - \frac{3}{4}N$ & 13.9 & 1.0\\
\hline
\textbf{singleQR} & $8NM^2 + \frac{7}{2}NM - 4M^3 -\frac{15}{4}M^2 - \frac{1}{4}M - \frac{1}{2}N$& 1.8 & 0.07\\
\hline
\end{tabular}
\end{table*}
\indent Noting that the problem of the single-QR algorithm is in the selection of the 
first antenna, we consider equally all antennas to be selected at the first iteration 
despite their condition. Thus, at first, the norms of the columns of the channel matrix are calculated, and their descending order \textbf{w} is obtained such that 
\begin{equation}
\left\|\textbf{h}_{w_1}\right\|^2 \geq \left\|\textbf{h}_{w_2}\right\|^2 
\geq \cdots \geq \left\|\textbf{h}_{w_N}\right\|^2. 
\end{equation}
At the first stage of the proposed maxR algorithm, the best metric $bm$ is initiated to $\infty$, the column with the largest power is considered at the first QRD iteration leading to reduction in $1/R_{1,1}^2$. The remaining ($N-1$) columns of the matrix \textbf{Q} are then orthogonalized and the one with largest power is selected for the second iteration. This process is repeated till obtaining the first antenna subset of $M$ elements. Also, the metric $D = \sum_{i=1}^M 1/R_{i,i}^2$ is calculated successively, and assigned to $bm$. At the second stage of the proposed algorithm, 
the $w_2$-th column of the matrix \textbf{H} is selected at the first iteration of the QRD, where the remaining columns are orthogonalized and the corresponding $D = 1/R_{1,1}^2$ is calculated. At any iteration, if $D$ exceeds the already-found best metric $bm$, the stage is stopped and the algorithm moves to the next stage. Otherwise, the algorithm proceeds until the last iteration, where $D$ is updated successively.\\
\indent Fig. \ref{maxAlg} gives the detailed description of the proposed maxR transmit antenna selection algorithm. Note that $\textbf{0}_n$ is the $n\times n$ matrix whose elements are all zeros, and $\mathcal{S}= \left\{1, 2, \cdots, N\right\}$ is the set of all available antennas.\\
\indent The proposed maxR algorithm is easily pipelined due to its parallel nature, where pipelining leads to tremendous reduction in the latency particularly for large $N$.
\section{Computational Complexity of the Transmit Antenna Selection Algorithms}
The computational complexities of the introduced transmit antenna selection algorithms are derived in this section. The addition and multiplication operations are counted as one flop and three flops, respectively. Also, the matrix inversion is obtained using the Gauss-Jordan row operations \cite{Arsham}. In Table \ref{table_complexity}, the computational complexities of the aforementioned algorithms are given for $K\,=\,M$. Moreover, two numerical examples for $(N = 8, M = 4)$ and $(N = 14, M = 10)$ are given, where $\text{C}/{\text{C}_{\text{opt}}}$ is the ratio between the complexity of the antenna selection algorithm and that of the optimum algorithm. Single-QR algorithm requires the least computational efforts among all the presented algorithms. The maximum computational complexity of proposed maxR algorithm is about 1$\%$ as compared to that of the optimum algorithm for $N = 14$ and $M = 10$. This maximum complexity of the proposed maxR is required when the algorithm is processed in parallel to reduce the latency, whereas complexity is reduced when the algorithm is processed sequentially.
\section{Simulation Results and Discussions}
\begin{figure}[t]
\centering
\includegraphics[width=7cm]{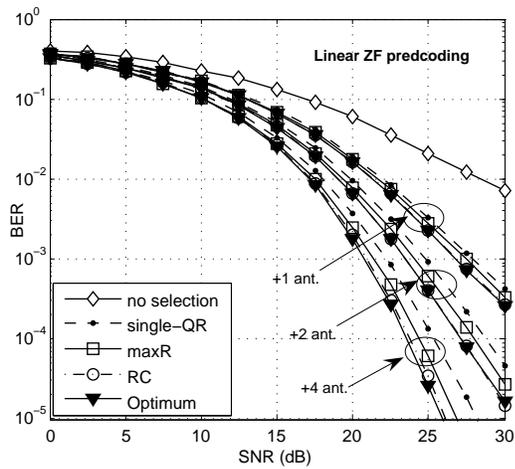}
\caption{Performance of the LZF precoding for several transmit antenna selection algorithms for $M = K = 4$, and $N = 5$, $6$, and $8$.}
\label{LZFber}
\end{figure}

\begin{figure}[t]
\centering
\includegraphics[width=7.25cm]{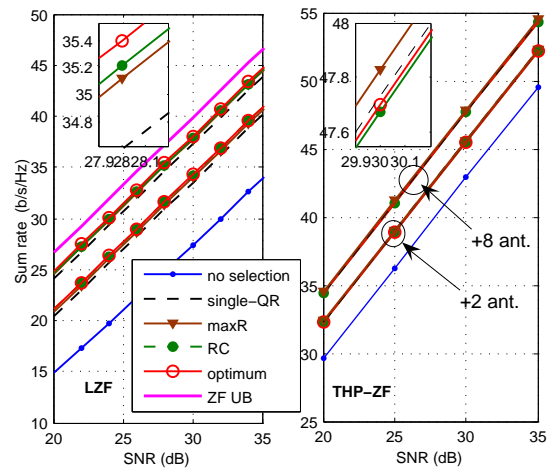}
\caption{Achieved sum rates by LZF (left) and THP-ZF (right) for $N = 6$ and $N = 12$.}
\label{Capacity}
\end{figure}

In this section we evaluate the bit error rate (BER) performance of the introduced antenna selection algorithms for LZF and THP-ZF precoding schemes using 16-QAM. The channel is considered to be perfectly known at the transmitter. We consider 4 uncooperative MSs each equipped with a single receive antenna, and a single BS equipped with $N = 5$, $6$, or $8$ transmit antennas from which only 4 are selected and switched to the 4 available RF chains.\\
\indent Fig. \ref{LZFber} depicts the BER performance of the LZF precoding for various transmit antenna selection algorithms. For $N = 8$, i.e., 4 additional antennas, and at BER of $10^{-4}$, degradations of 0.25dB, 0.8dB, and 1.8dB are remarked when RC, maxR, and single-QR antenna selection algorithms are used, respectively, compared to the optimum performance.\\
\begin{figure}[t]
\centering
\includegraphics[width=7cm]{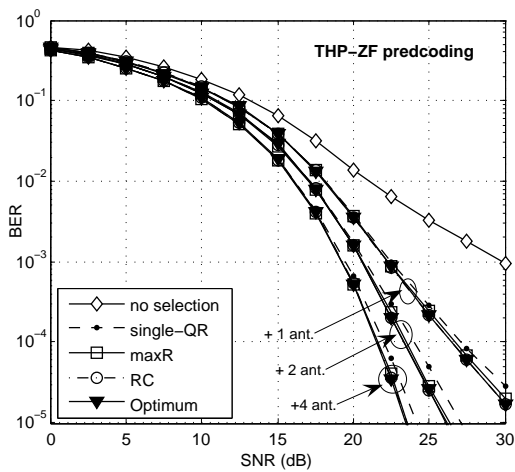}
\caption{Performance of THP-ZF for several transmit antenna selection algorithms for $M = K = 4$, and $N = 5$, $6$, and $8$.}
\label{THPZFber}
\end{figure}
\indent Fig. \ref{THPZFber} depicts the BER performance of the THP-ZF scheme for various transmit antenna selection algorithms. For $N = 8$ and at a target BER of $10^{-4}$, RC, maxR, and single-QR antenna selection algorithms lead to degradations of 0.05dB, 0.09dB, and 0.5dB, respectively, compared to the optimum performance.\\
\indent From Fig. \ref{LZFber} and \ref{THPZFber}, we conclude that the linear precoding scheme is more sensitive to the antenna selection scheme compared to the THP algorithm for which a quasi-optimum performance is achieved using sub-optimal antenna selection algorithms. Moreover, THP-ZF outperforms LZF precoding for all antenna configurations and for the different antenna selection algorithms. For instance, for $N = 8$ and at BER of $10^{-5}$, THP-ZF outperforms LZF by 2.65dB.\\
\indent Fig. \ref{Capacity} shows the achieved sum rates for both LZF and THP-ZF precoding schemes for $N = 6$ and $N = 12$, i.e., 2 and 8 additional antennas, respectively. The sum rate achieved by LZF precoding is remarkably increased for the few first additional antennas. This is due to its sensitivity to the $\gamma$ factor which is reduced as the number of additional antennas is increased. For large number of additional antennas, the $\text{E}[\gamma] \approx 1$ and the upper-bound on the capacity can be achieved. At SNR of $30$dB, The SNR-optimum algorithm achieves the best sum rate, while RC, the proposed maxR, and single-QR algorithms lag the optimum capacity by 0.045, 0.07, and 0.2 bits/s/Hz/user, respectively.\\
\indent In case of THP-ZF, the proposed max-R algorithm achieves the best sum rate followed by single-QR and the optimum scheme, while RC algorithm achieves the lowest sum rate capacity. This is because the sum rate of THP-ZF is directly affected by the diagonal elements of the matrix \textbf{R} which are maximized in both single-QR and the proposed maxR algorithms.


\section{Conclusions}
In this paper, we introduced four transmit antenna selection algorithms for downlink MU-MIMO systems, where each user is equipped with a single receive antenna. The optimal and a greedy suboptimal transmit antenna selection algorithms are presented. Furthermore, the suboptimal QRD-based antenna selection is investigated, then we introduce the conventional QRD-based algorithm and propose an efficient QRD-based transmit antenna scheme (maxR), which is both implementation and performance efficient. Simulation results show that the proposed maxR antenna selection algorithm performs close to the optimal algorithm while requiring only a small fraction of its computational complexity.





\begin{thebibliography}{1}
\bibitem{Costa}
M. Costa, ``Writing on dirty paper,'' {\em IEEE Transactions on Information Theory,} vol. IT-29, pp. 439–441, May 1983.

\bibitem{Peel}
C. Peel, B. Hochwald, and L. Swindlehurst, ``A vector-perturbation technique for near-capacity multiantenna multiuser communication - part I: channel inversion and regularization,'' {\em IEEE Transactions on Communications,} vol. 53, no. 1, pp. 195-202, January 2005.

\bibitem{Tomlinson}
M. Tomlinson, ``New automatic equalizer employing modulo arithmetic,''
{\em Electronics Letters,} vol. 7, pp. 138–139, Mar. 1971.

\bibitem{Harashima}
H. Harashima and H. Miyakawa, ``Matched-transmission technique for
channels with intersymbol interference,'' {\em IEEE Transactions on Communications,} vol.
no. 20, pp. 774–780, Aug. 1972.

\bibitem{Gucluoglu}
T. Gucluoglu, and T. Duman, ``Performance analysis of transmit and receive antenna selection over flat fading channels,'' {\em IEEE Transactions on Wireless Communications,} vol. 7, no. 8, pp. 3056-3055, August 2008.

\bibitem{Molisch}
A. Molisch ``MIMO systems with antenna selection - an overview,'' in {\em Proceedings of the Radio and Wireless Conference,} August 2003, pp. 167-170.

\bibitem{Sanayei2}
S. Sanayei, and A. Nosratinia, ``Antenna selection in MIMO systems,'' {\em IEEE Communications Magazine,} vol. 42, issue 10, pp. 68-73, October 2004.

\bibitem{Chen}
R. Chen, J. Andrews, and R. Heath, Jr., ``Efficient transmit antenna selection for multiuser MIMO systems with block diagonalization'' in {\em Proceedings of the IEEE Global Telecommunications Conference,} Nov. 2007, pp. 3499-3503.


\bibitem{Arsham}
H. Arsham and M. Oblak, ``Matrix inversion: a computational algebra approach,'' {\em International Journal of Mathematical Education in Science and Technology,} vol. 27, no. 4, pp. 599-605, 1996.

\end{thebibliography}
%

\end{document}